\documentclass[aps,prl,reprint]{revtex4-1}
\usepackage{amsmath}
\usepackage{graphicx}
\graphicspath{{figures/}}
\usepackage[separate-uncertainty = true,multi-part-units=single]{siunitx}
\usepackage{hyperref}

\renewcommand{\vec}[1]{\mathbf{#1}}
\newcommand{\abs}[1]{\left\vert #1 \right\vert}

\begin{document}

\title[Scanning Holographic Camera]{Scanning Camera for Continuous-Wave Acoustic Holography}

\author{Hillary W. Gao}
\author{Kimberly I. Mishra}
\author{Annemarie Winters}
\author{Sidney Wolin}
\author{David G. Grier}
 \email{david.grier@nyu.edu}
\affiliation{Department of Physics and Center for Soft Matter Research,
New York University, New York, NY 10003, USA
}

\date{\today}

\begin{abstract}
We present a system for measuring the 
amplitude and phase profiles of the pressure field of
a harmonic acoustic wave with the goal of 
reconstructing the volumetric sound field.
Unlike optical holograms that cannot be reconstructed
exactly because of the inverse problem, 
acoustic holograms are completely
specified in the recording plane.
We demonstrate volumetric reconstructions of
simple arrangements of objects using
the Rayleigh-Sommerfeld diffraction integral,
and introduce a technique to analyze the
dynamic properties of insonated objects.
\end{abstract}

\maketitle

Most technologies for acoustic imaging
use the temporal and spectral characteristics
of acoustic pulses to map interfaces
between distinct phases.
This is the basis for sonar \cite{hayes2009synthetic}, 
and medical and industrial ultrasonography \cite{fenster1996}.
Imaging continuous-wave sound fields
is useful for industrial and environmental
noise analysis, particularly for
source localization \cite{sheng2005maximum}.
Substantially less attention has been paid to 
visualizing the amplitude and phase
profiles of sound fields for their own
sakes, with most effort being focused on
visualizing the near-field acoustic
radiation emitted by localized sources,
a technique known as near-field
acoustic holography (NAH) \cite{maynard1985nearfield,veronesi1987nearfield,williams1999fourier}.
The advent of acoustic manipulation
in holographically structured sound fields 
\cite{marston06,zhang2014generation,marzo15,melde2016holograms,tian2017acoustic}
creates a need for effective
sound-field visualization.
Here, we demonstrate a scanning
acoustic camera that combines lockin detection
with a polargraph for flexible large-area scanning
to accurately
record the wavefront structure
of acoustic travelling waves.
Borrowing techniques from optical holography,
we use Rayleigh-Sommerfeld back-propagation
\cite{lee07}
to reconstruct the three-dimensional 
sound field associated with the complex
pressure field in the measurement plane.
These reconstructions, in turn, provide
insights into the dynamical properties
of objects immersed in the acoustic field.

\section{Holography}
\label{sec:holography}

A harmonic traveling wave at frequency $\omega$
can be described by a complex-valued wave function,
\begin{equation}
  \label{eq:wave}
  \psi(\vec{r},t)
  =
  u(\vec{r}) \, e^{i \varphi(\vec{r})}
  e^{-i \omega t},
\end{equation}
that is characterized by real-valued
amplitude and phase profiles, $u(\vec{r})$
and $\varphi(\vec{r})$, respectively.
Eq.~\eqref{eq:wave} can be generalized
for vector fields by incorporating separate
amplitude and phase profiles for each of the
Cartesian coordinates.
The field propagates
according to the wave equation,
\begin{equation}
  \label{eq:waveequation}
  \nabla^2 \psi = k^2 \psi ,
\end{equation}
where the wave number $k$ is the magnitude
of the local wave vector,
\begin{equation}
  \label{eq:wavevector}
  \vec{k}(\vec{r}) 
  =
  \nabla \varphi(\vec{r}).
\end{equation}

A hologram is produced by illuminating an object
with an incident wave, $\psi_0(\vec{r},t)$, whose 
amplitude and phase profiles are $u_0(\vec{r})$ and
$\varphi_0(\vec{r})$, respectively.  The
object scatters some of that wave to produce
$\psi_s(\vec{r},t)$, which propagates to the imaging
plane.
In-line holography uses the remainder of
the incident field as a reference wave that
interferes with the scattered field to produce
a superposition
\begin{equation}
  \label{eq:superposition}
  \psi(\vec{r},t) = \psi_0(\vec{r},t) + \psi_s(\vec{r},t)
\end{equation}
whose properties are recorded.
The wave equation then can be used to numerically
reconstruct the three-dimensional field from
its value in the plane.
In this way, numerical back-propagation can
provide information about the object's position relative to
the recording plane as well as its size, shape and
properties.
The nature of the recording determines how much
information can be recovered.

\subsection{Optical holography: Intensity holograms}
\label{sec:optical}

Optical cameras record the intensity of the
field in the plane, and so discard all of the
information about the wave's direction of
propagation that is encoded in the phase.
Interfering the scattered wave with a reference field
yields an intensity distribution,
\begin{subequations}
\label{eq:intensity}
\begin{align}
  I(\vec{r}) 
  & =
    \abs{\psi(\vec{r},t)}^2 \\
  & =
    \abs{
    u_0(\vec{r}) \, e^{i\varphi_0(\vec{r})}
    +
    u_s(\vec{r}) \, e^{i\varphi_s(\vec{r})}}^2,
\end{align}
\end{subequations}
that blends information about both the
amplitude and the phase of the scattered wave
into a single scalar field.

The properties of the scattered field can be
interpreted most easily
if the incident field can be modeled as a
unit-amplitude plane
wave,
\begin{subequations}
\label{eq:scatteringmodel}
\begin{equation}
  \psi_0(\vec{r},t) \approx e^{ikz} e^{-i\omega t},
\end{equation}
in which case,
\begin{equation}
  I(\vec{r}) 
  \approx
  \abs{1 + u_s(\vec{r}) \, e^{i \varphi_s(\vec{r})}}^2.
\end{equation}
If, furthermore, the scattering process may be
modeled with a transfer function,
\begin{equation}
  \psi_s(\vec{r}) = T(\vec{r} - \vec{r}_s) \, \psi_0(\vec{r}_s),
\end{equation}
\end{subequations}
then $I(\vec{r})$ can be used to estimate parameters
of $T(\vec{r})$, including the position and properties
of the scatterer.

This model has proved useful for
interpreting in-line 
holograms of micrometer-scale colloidal particles
\cite{lee07a}.
Fitting to Eq.~\eqref{eq:scatteringmodel} 
can locate a colloidal sphere in three dimensions
with nanometer precision \cite{fung11,krishnatreya14a}.
The same fit yields estimates for the 
sphere's diameter and refractive index
to within a part per thousand \cite{krishnatreya14a}.
Generalizations of this method \cite{wang14using}
work comparably well for tracking clusters 
of particles
\cite{fung12,perry12,fung13}.

\subsection{Rayleigh-Sommerfeld back propagation}
\label{sec:rayleighsommerfeld}

The success of fitting methods is based on
\emph{a priori} knowledge of the nature of
the scatterer, which is encoded in the
transfer function, $T(\vec{r})$.
In instances where such knowledge is not 
available, optical holograms also can be used
as a basis for reconstructing the scattered
field, $\psi_s(\vec{r},t)$, in three dimensions.
This reconstruction can serve as a proxy for
the structure of the sample.

One particularly effective reconstruction
method \cite{lee07,cheong10} is based on the Rayleigh-Sommerfeld
diffraction integral \cite{goodman05}.
The field, $\psi(x, y, 0,t)$ in the
imaging plane, $z = 0$, propagates to
point $\vec{r}$ in
plane $z$ as \cite{goodman05}
\begin{subequations}
\label{eq:rsconvolution}
\begin{equation}
  \psi(\vec{r},t)
  =
  \psi(x, y, 0, t) \otimes h_z(x, y),
\end{equation}
where
\begin{equation}
  h_z(x, y) =
  \frac{1}{2\pi} \, \frac{d}{dz} 
  \frac{e^{i kr}}{r}.
\end{equation}
\end{subequations}
is the Rayleigh-Sommerfeld propagator.
The convolution in Eq.~\eqref{eq:rsconvolution} is
most easily computed with the Fourier convolution
theorem using the Fourier transform of the propagator,
\begin{equation}
  \label{repropagator}
  H_z(\vec{q}) = e^{-i z \sqrt{k^2 - q^2}}.
\end{equation}
Equation~\eqref{eq:rsconvolution} can be used
to numerically propagate a measured wave
back to its source, thereby reconstructing
the three-dimensional field responsible for the
recorded pattern.

\begin{figure}[t!]
\includegraphics[width=0.9\columnwidth]{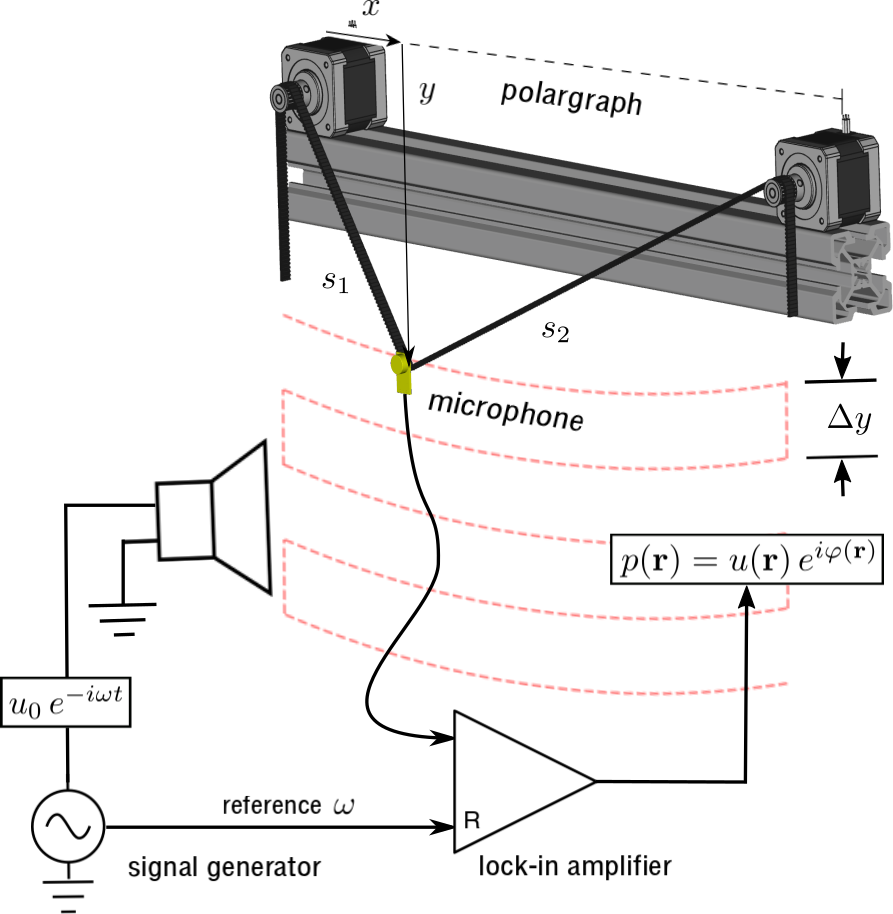}
\caption{(color online) 
Schematic representation of the scanning acoustic camera.
A polargraph composed of two stepping motors and a timing 
belt translates a microphone across the field of view
in a serpentine pattern.
A harmonic sound field is created by an audio speaker
driven by a signal generator.
The signal detected by the microphone is analyzed by a
lock-in amplifier to obtain the amplitude $u(\vec{r})$
and phase $\varphi(\vec{r})$
of the sound's pressure field at each position 
$\vec{r} = (x, y)$.}
\label{fig:schematic}
\end{figure}

\subsection{Acoustic holography: Complex holograms}
\label{sec:acousticholography}

Early implementations of acoustic holography
resembled optical holography in recording the
intensity of the sound field \cite{mueller1971acoustic}.
Sound waves of modest intensity, however,
are fully characterized by a scalar pressure field,
$p(\vec{r},t)$, whose amplitude and phase 
can be measured directly.
In that case, Rayleigh-Sommerfeld back-propagation
can be used to reconstruct the
complex sound field with an accuracy that
is limited by instrumental noise and by
the size of the recording area.
For appropriate systems, the transfer-function
model can be used to obtain information
about scatterers in the field.

Whereas optical holography benefits from
highly developed camera technology,
implementations of quantitative acoustic
holography must confront a lack of suitable
area detectors for sound waves.
Commercial acoustic transducer arrays typically
include no more than a few dozen elements.
Conventional area scanners yield excellent results over small areas
\cite{melde2016holograms}
but become prohibitively expensive for large-area scans.
We therefore introduce flexible and 
cost-effective techniques for
recording complex-valued acoustic holograms.

\section{Scanning acoustic camera}
\label{sec:instrument}

Figure~\ref{fig:schematic} depicts our implementation of a scanning
acoustic camera for holographic imaging.
A signal generator 
(Stanford Research Systems DS345)
drives an audio speaker at a desired frequency $\omega$.
The resulting sound wave propagates to a microphone
whose output is analyzed
by a dual-phase lock-in amplifier 
(Stanford Research Systems SR830) 
referenced to the signal generator.
Our reference implementation operates at
\SI{8}{\kilo\hertz}, which corresponds to
a wavelength of \SI{42.5}{\mm}.

The lock-in amplifier records the amplitude and phase
of the pressure field at the microphone's position.
We translate the microphone
across the $x$-$y$ plane using
a flexible low-cost two-dimensional scanner known as a polargraph
that initially was 
developed for art installations \cite{lehni02}.
Correlating the output of the lock-in amplifier with the position of the
polargraph yields a map of the complex pressure field in the plane.

\subsection{Polargraph for flexible wide-area scanning}
\label{sec:polargraph}

The polargraph consists of two stepper motors 
(Nema 17, 200 steps/rev)
with toothed pulleys (16 tooth, GT2, \SI{7.5}{\mm} diameter)
that control the movement of a flexible GT2 timing belt,
as indicated in Fig.~\ref{fig:schematic}.
The acoustic camera's 
microphone (KY0030 high-sensitivity sound
module) is mounted on a laser-cut
support that hangs under gravity from the
middle of the timing belt.
The motors' rotations determine 
the lengths, $s_1(t)$ and $s_2(t)$ 
of the two chords of
timing belt at time $t$,
and therefore the position of the microphone,
$\vec{r}(t)$.
If the motors are separated by distance $L$
and the microphone initially is located
at height $y_0$ below their midpoint, then
\begin{subequations}
\label{eq:position}
\begin{align}
  \vec{r}(t)
  & =
    x(t) \, \hat{x} + y(t) \, \hat{y}, \quad\text{where} \\
  x(t) & = \frac{s_1^2 - s_2^2}{2L} \quad \text{and}\\
  y(t) & = \sqrt{\frac{s_1^2+s_2^2}{2}-\frac{L^2}{4}-x^2(t)}-y_0. 
\end{align}
\end{subequations}
In our implementation, $L = \SI{1}{\meter}$
and $y_0 = \SI{10}{\cm}$.

The stepper motors are
controlled with an Arduino microcontroller
that is addressed by software running on a
conventional computer \footnote{The open-source 
software for this instrument is available online
at \url{http://github.com/davidgrier/acam}}.
Smoothest operation is obtained by running the stepper
motors at constant rates.  This results in the
microphone translating through a serpentine pattern,
as indicated in Fig.~\ref{fig:schematic}.
Horizontal sweeps of \SI{0.6}{\meter} are separated
by vertical steps of $\Delta y = \SI{5}{\mm}$,
and the motors' step rates are configured to maintain
a constant scan speed of \SI{46.5}{\mm\per\second}.

The polargraph is deployed by mounting the
stepper motors at the upper corners of the
area to be scanned.
Our implementation has the motors
mounted on a frame constructed from
extruded 1-inch square aluminum T-slot stock,
as indicated in Fig.~\ref{fig:schematic}.
The scan area is limited by the length of
the timing belt and by the requirement that
both chords of the belt remain taut throughout
the scan.  This can be facilitated by adding
weight to the microphone's mounting bracket
or by adding a tensioning cable.
Mechanical vibrations during the scan are
substantially smaller than the
effective pixel size of the resulting 
acoustic holograms.

\subsection{Lockin detection}
\label{sec:lockin}

While the polargraph scans the microphone
across the field view, the lockin amplifier
reports the amplitude and phase of the 
signal detected by the microphone.
Setting the lockin's time constant 
to \SI{30}{\ms} effectively suppresses
background noise yet is short enough to
enable independent measurements at
\SI{0.1}{\second} intervals.
Given the translation speed, this
corresponds to an effective
spatial resolution of 
\SI{4.65 \pm 0.05}{\mm}.
The vertical separation, $\Delta y$,
between horizontal sweeps is set
accordingly.
We report the relative amplitude of the
instrument's response with full scale
corresponding to roughly \SI{1}{\pascal}.

The amplifier's measurements
of $u(t)$ and $\varphi(t)$ at time $t$
are associated with the polargraph's
position, $\vec{r}(t)$, that is computed
with Eq.~\eqref{eq:position} at the same time.
Because the lockin amplifier's readout
is not synchronized to the polargraph's
motion, this yields an irregularly gridded
representation of the complex pressure field.
We therefore resample
$u(t)$ and $\varphi(t)$ onto a
\SI{128 x 128}{pixel} Cartesian grid 
(\num{16384} effective pixels)
with \SI{4.90 \pm 0.08}{\milli\meter} spacing using bilinear interpolation.
This measurement grid is finer
than the wavelength of sound
and is substantially larger than can be achieved with
currently available microphone arrays.
The parameters selected here yield one 
complete measurement in about \SI{25}{\minute}.

\begin{figure*}[t!]
\centering
\includegraphics[width=\textwidth]{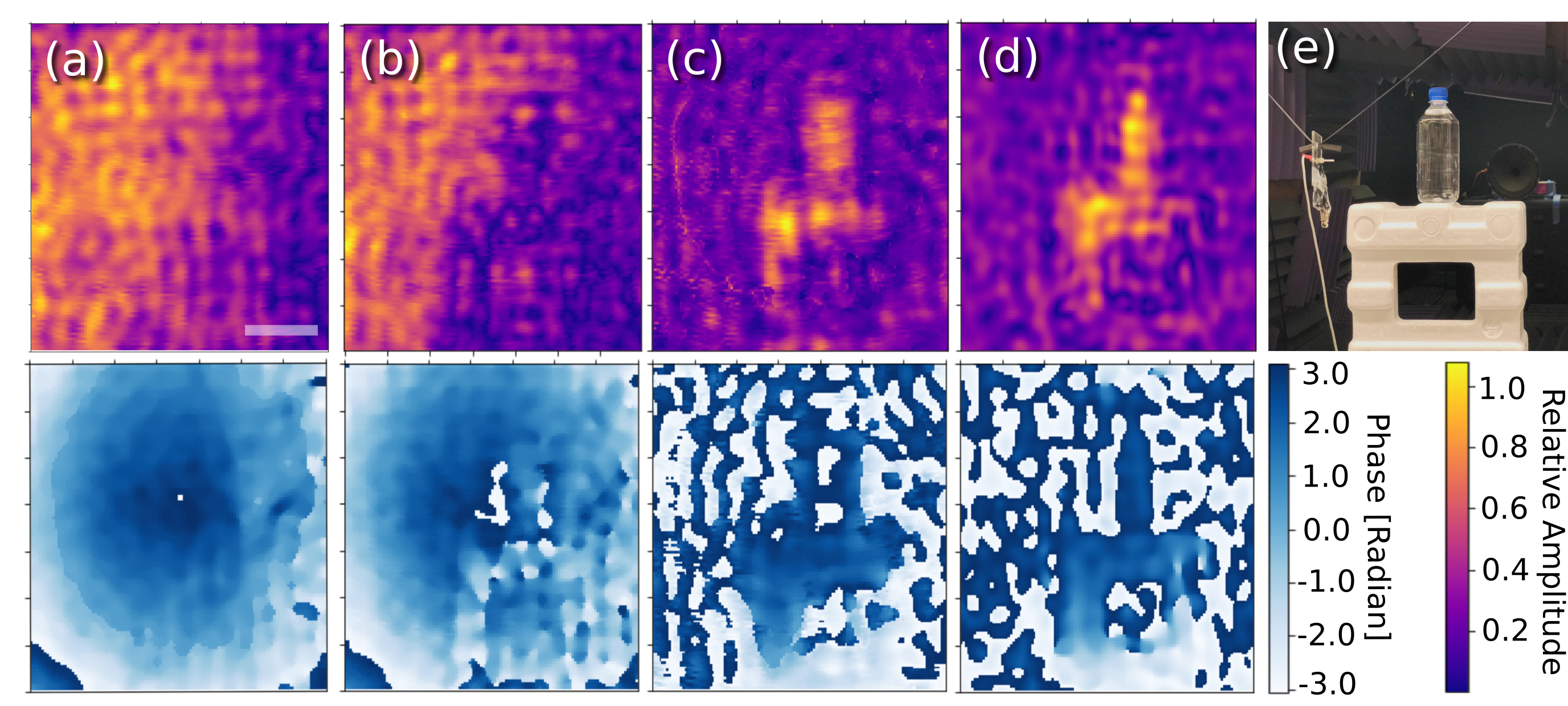}
\caption{(color online)
  Acoustic holography of objects in a sound field.
  (a) Amplitude (top) and phase (bottom) of the harmonic
  field at \SI{8}{\kilo\hertz} projected by a speaker
  nearly \SI{2}{\meter} from the imaging plane.
  Scale bar corresponds to \SI{15}{\cm}.
  (b) The same field of view with scattering objects
  in the foreground.  (c) Background-subtracted estimate
  for the scattered field in the imaging plane.
  (d) Refocused acoustic image obtained from (c) by
  Rayleigh-Sommerfeld back-propagation with 
  Eq.~\eqref{eq:rsconvolution}.
  (e) Photograph of the scene recorded by the acoustic camera,
  including the camera's microphone.}
\label{fig:reconstruction}
\end{figure*}

\section{Results}
\label{sec:results}

\subsection{Holographic imaging}
\label{sec:imaging}

Figure~\ref{fig:reconstruction}(a) shows the amplitude
and phase of the sound field reaching the recording plane
when the speaker is located nearly \SI{2}{\meter}
away.
The recording plane
is normal to the direction of sound propagation and the walls of
the \SI{1.5 x 1.5}{\meter} experimental volume are lined with
2-inch wedge acoustic tiles to minimize reflections.
Even so, off-axis reflections reach the measurement plane
and interfere with the directly propagating sound field
in the measurement plane.  These interference features
are particularly evident in the amplitude profile in
Fig.~\ref{fig:reconstruction}(a)
and appear in all of the results that we present.

The phase profile projected by the speaker is smoothly
curved, which is expected for the
diverging pressure field from a localized source.
Taking the speed of sound in air to be
$v = \SI{340}{\meter\per\second}$,
the center of 
curvature of the phase profile
is located \SI{160(10)}{\cm} away from the
observation plane, which is consistent with the
the speaker's position.
Because the lockin amplifier measures
phase delay over a limited range, we
present the phase modulo
$2\pi$.
This causes abrupt transitions 
to appear in the rendering.
We move these out of the field of view by
shifting the lockin amplifier's phase response by
$\Delta \varphi = \SI{0.02}{\degree}$.

\begin{figure}
\includegraphics[width=0.8\columnwidth]{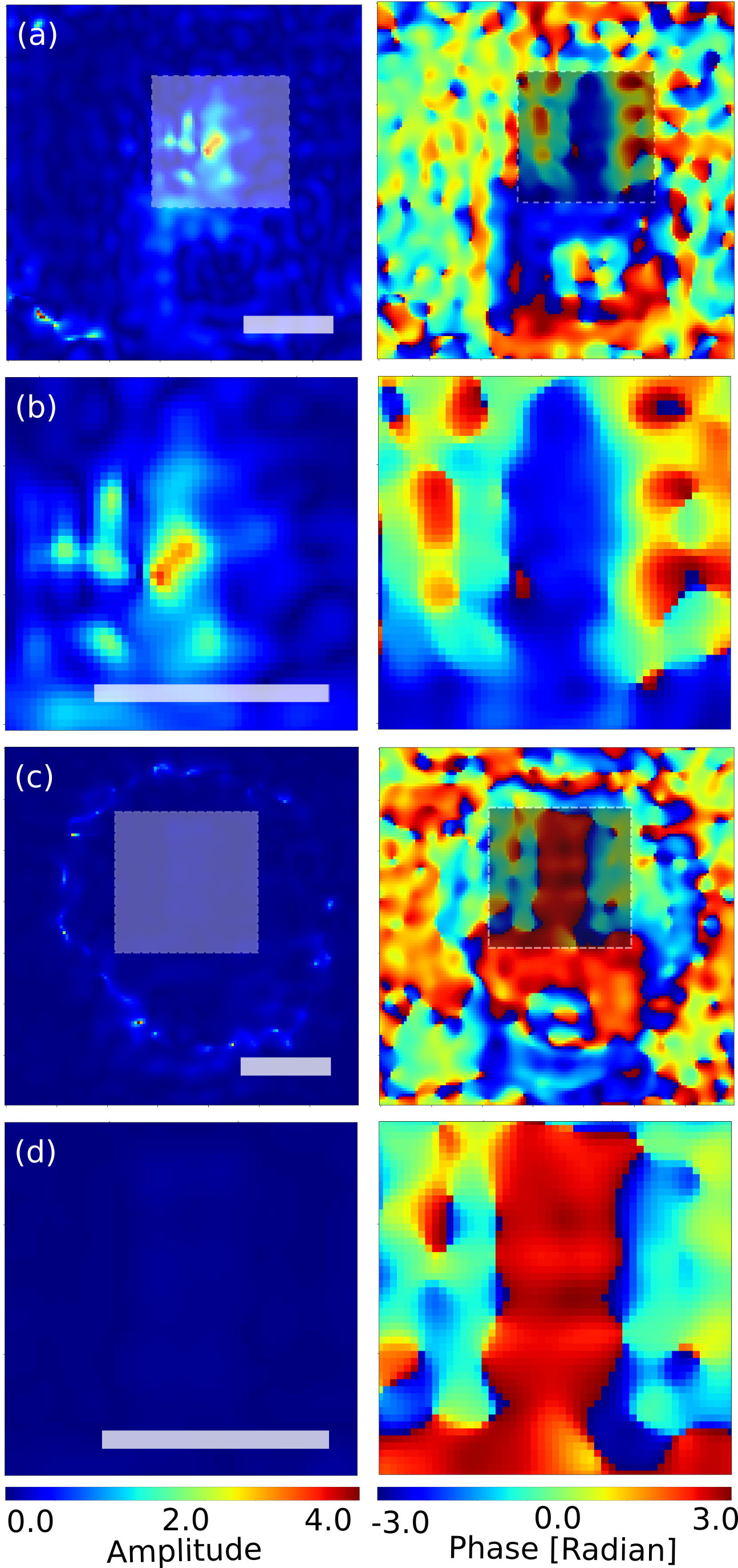}
\caption{(color online)
  Detecting and localizing mechanical resonances in insonated objects.
  (a) Amplitude (left) and phase (right) 
  of the transfer function for the system in 
  Fig.~\ref{fig:reconstruction}.  Shaded boxes indicate the
  region of interest that is presented in (b). (c) Amplitude
  and phase of the transfer function for a
  system identical except with a similar-sized block of wood
  instead of a water bottle. Shaded boxes indicate the
  region of interest that is presented in (d).
  Scale bars correspond to \SI{15}{\cm}.}
\label{fig:normalized}
\end{figure}

The complex pressure field, $p_0(\vec{r})$, serves
as the background field for other holograms recorded by this
instrument.
Figure~\ref{fig:reconstruction}(b) shows the same field of view
partially obstructed by a collection of objects, specifically
a plastic bottle filled with water placed atop a styrofoam box.
The objects appear as a shadow in the amplitude profile
and as a pattern of discontinuities in the phase profile.
Because the phase profile is wrapped into the range 
$0 \leq \varphi(\vec{r}) \leq 2 \pi$, it
cannot be used to measure the speed of sound within the
objects.
It does, however, provide enough information to
reconstruct the three-dimensional sound field.

The complex pressure field in the imaging plane,
$p(\vec{r}) = p_0(\vec{r}) + p_s(\vec{r})$,
includes both the source field, $p_0(\vec{r})$,
and the field scattered by the objects, $p_s(\vec{r})$.
The difference $\Delta p(\vec{r}) = p(\vec{r}) - p_0(\vec{r})$
between the recorded hologram of the object and 
the previously recorded background
pressure field is an estimate of the scattered field
due to the object in the measurement plane.
This is plotted in Fig.~\ref{fig:reconstruction}(c).
The object's shadow appears bright in the amplitude
distribution because it differs substantially from the
background amplitude.
The sign of this difference 
is encoded in the phase,
which confirms that
the object has reduced the sound level directly
downstream.

Figure~\ref{fig:reconstruction}(d) shows the result
of numerically reconstructing the field at a
distance $z = \SI{15.6}{\cm}$ 
behind the recording plane.
This effectively brings the object into focus without
otherwise distorting its image.  The result
is consistent with a photograph of the scene,
which is reproduced in Fig.~\ref{fig:reconstruction}(e).

\subsection{Holographic characterization of dynamical properties}
\label{sec:dynamics}

If the object scattering the incident wave is not substantially 
larger than the wavelength, the wave it scatters may
be modeled as the incident wave in the scattering plane,
$z = z_s$, modified by a complex transfer function:
\begin{equation}
    \label{eq:transferfunction}
    p_s(x,y,z_s) = T(\vec{r}) \, p_0(x,y,z_s).
\end{equation}
We estimate the transfer function by numerically back-propagating
the incident wave to the scattering plane using Eq.~\eqref{eq:rsconvolution},
and using it to normalize the back-propagated scattered wave:
\begin{equation}
    \label{eq:scattering}
    T(\vec{r}) = \frac{p(x,y,z_s)}{p_0(x,y,z_s)} - 1.
\end{equation}
In the absence of any
object, we expect
$T(\vec{r}) = 0$.

Figure \ref{fig:normalized}(a) shows the amplitude
and phase of the transfer function, $T(\vec{r})$,
of the sample from Fig.~\ref{fig:reconstruction}.
As expected, the background amplitude of the
computed transfer function has
magnitude and phase near zero.
The transfer function of a passive object should
advance the phase of the incident wave while
reducing its amplitude.
We therefore expect $\abs{T(\vec{r})} \leq 1$,
with $T(\vec{r}) = -1$ corresponding to a
perfect absorber.
In fact, the water bottle's
transfer function presents a well localized
peak of magnitude \num{5} near one corner
of the object.
This can be seen more clearly in the expanded
field of view in Fig.~\ref{fig:normalized}(b).
The peak in the transfer function's amplitude
is associated with a 
localized reversal of its phase.
Most of the bottle's acoustic transfer function
has phase $\arg T(\vec{r}) \approx - \pi$, which
suggests that it is removing energy from the field
in those regions.  The abrupt transition through
$\arg T(\vec{r}) = 0$ to $\arg T(\vec{r}) = \pi$
near the bottle's left side is consistent with
localized phase reversals in the region of the
peak in $\abs{T(\vec{r})}$.
These observations suggest that this feature
may be identified with a resonant mode of the water bottle
that focuses acoustic energy.

To test this interpretation, we replace the
water bottle with a comparably sized block of
wood and repeat the measurement.
The resulting transfer function is plotted
in Fig.~\ref{fig:normalized}(c), with an
expanded field of view in Fig.~\ref{fig:normalized}(d).
The magnitude of the block's transfer function is
no greater than \num{0.2} across the entire field
of view in Fig.~\ref{fig:normalized}(d).
The phase profile within the block similarly
lacks any prominent features.

\section{Conclusions}
\label{sec:conclusions}

We have presented a scanning acoustic camera that can be
deployed flexibly to record the amplitude and phase
profiles of harmonic pressure fields over wide areas.
We further have demonstrated the use of the
Rayleigh-Sommerfeld propagator to back-propagate
the measured acoustic hologram to reconstruct
the three-dimensional sound field.
This is useful for numerically refocusing a recorded
image of an object immersed in the sound field.
We have shown, furthermore, that
numerical back propagation can be used to estimate 
an object's complex acoustic transfer function
and thus to probe its dynamical properties at the
driving frequency of the sound field.
We demonstrate this by imaging a resonance in a
container of water.

We anticipate that the acoustic imaging
capabilities we have described will be useful for
research groups deploying structured acoustic
fields for communication, sensing and 
manipulation.
In such cases, the scanning technique provides
a cost-effective means to characterize the projected
wave, even when it covers a very large area.
The camera also is useful for creating
images of scenes
in continuous harmonic waves.
Our implementation is deployed for 
imaging in transmission.
Reflected and oblique imaging
also should be possible.
Numerically refocused acoustic imaging 
is particularly useful for 
remote imaging of dynamical properties.

While our implementation is based on lockin
detection of harmonic waves, generalizations
to broad-spectrum sources can be implemented
with correlation-based detection.
The scanned approach similarly should lend
itself to imaging in noise\cite{potter1994acoustic},
including imaging
of dynamical properties.

\begin{acknowledgments}
This work was supported by the MRSEC program of the
National Science Foundation through award number
DMR-1420073.
\end{acknowledgments}


\end{document}